\documentclass[aps,prl,twocolumn,showpacs,superscriptaddress]{revtex4}

\usepackage[dvips]{graphicx} %[pdftex][dvips]
\usepackage{natbib}

\newcommand{\festx}{FeSe$_{x}$Te$_{1-x}$}
\newcommand{\fest}{FeSe$_{0.42}$Te$_{0.58}$}

\begin{document}
% Use the \preprint command to place your local institutional report
% number in the upper righthand corner of the title page in preprint mode.
% Multiple \preprint commands are allowed.
% Use the 'preprintnumbers' class option to override journal defaults
% to display numbers if necessary
%\preprint{}

%Title of paper
\title{Strong electron correlations in the normal state of FeSe$_{0.42}$Te$_{0.58}$}
%\title{Correlated electronic structure of \fest{} from ARPES}
%\title{Strong and orbital dependent mass enhancement in the normal state of \fest}
% the 'origin' in this title only refers to band-structure vs renormalization

% repeat the \author .. \affiliation  etc. as needed
% \email, \thanks, \homepage, \altaffiliation all apply to the current
% author. Explanatory text should go in the []'s, actual e-mail
% address or url should go in the {}'s for \email and \homepage.
% Please use the appropriate macro foreach each type of information

% \affiliation command applies to all authors since the last
% \affiliation command. The \affiliation command should follow the
% other information
% \affiliation can be followed by \email, \homepage, \thanks as well.
\author{A. Tamai}
\affiliation{School of Physics and Astronomy, University of St. Andrews, St. Andrews, Fife KY16 9SS, United Kingdom}
\author{A.Y. Ganin}
\affiliation{Department of Chemistry, University of Liverpool, Liverpool L69 7ZD, United Kingdom}
\author{E. Rozbicki}
\affiliation{School of Physics and Astronomy, University of St. Andrews, St. Andrews, Fife KY16 9SS, United Kingdom}
\author{J. Bacsa}
\affiliation{Department of Chemistry, University of Liverpool, Liverpool L69 7ZD, United Kingdom}
\author{W. Meevasana}
\affiliation{School of Physics and Astronomy, University of St. Andrews, St. Andrews, Fife KY16 9SS, United Kingdom}
\author{P.D.C. King}
\affiliation{School of Physics and Astronomy, University of St. Andrews, St. Andrews, Fife KY16 9SS, United Kingdom}
\author{M. Caffio}
\affiliation{School of Chemistry, University of St. Andrews, St. Andrews, Fife KY16 9ST, United Kingdom}
\author{R. Schaub}
\affiliation{School of Chemistry, University of St. Andrews, St. Andrews, Fife KY16 9ST, United Kingdom}
\author{S. Margadonna}
\affiliation{School of Chemistry, University of Edinburgh, Edinburgh EH9 3JJ, United Kingdom }
\author{K. Prassides}
\affiliation{Department of Chemistry, Durham University, Durham DH1 3LE, United Kingdom }
\author{M.J. Rosseinsky}
\affiliation{Department of Chemistry, University of Liverpool, Liverpool L69 7ZD, United Kingdom}
\author{F. Baumberger}
\affiliation{School of Physics and Astronomy, University of St. Andrews, St. Andrews, Fife KY16 9SS, United Kingdom}
%\email[]{}

%\homepage[]{Your web page}
%\thanks{}
%\altaffiliation{}

\date{\today}

\begin{abstract}
We investigate the normal state of the '11' iron-based superconductor FeSe$_{0.42}$Te$_{0.58}$ by angle resolved photoemission. Our data reveal a highly renormalized quasiparticle dispersion characteristic of a strongly correlated metal. We find sheet dependent effective carrier masses between $\approx 3 - 16$~m$_{e}$ corresponding to a mass enhancement over band structure values of $m^{*}/m_{\rm{band}}\approx 6 - 20$. This is nearly an order of magnitude higher than the renormalization reported previously for iron-arsenide superconductors of the '1111' and '122' families but fully consistent with the bulk specific heat. 
\end{abstract}

% insert suggested PACS numbers in braces on next line
% insert suggested keywords - APS authors don't need to do this
%\keywords{}

%\maketitle must follow title, authors, abstract, \pacs, and \keywords
\maketitle

% introduction
The discovery of superconductivity with a high transition temperature in layered ferro-pnicitides \cite{kam06ea,ren08ea} was a major advance in material science and has motivated intense work on ferrous superconductors. Although the electronic structure of the ferro-pnicitides is clearly different from the cuprate high-temperature superconductors, they share a tendency towards magnetically ordered states in the vicinity of the superconducting phase. A key-difference to cuprates, however, is the absence of a Mott--insulating phase. This raises the question of whether electronic correlations are important in iron based superconductors and how they manifest themselves in the electronic structure. 

Early dynamical mean field theory (DMFT) calculations of the electronic correlations reported conflicting results ranging from weakly enhanced metallic states to strongly correlated systems on the verge of a Mott metal-insulator transition \cite{hau08,ani09}. 
However, spectroscopic measurements \cite{lu08ea,zab09ea,yi09ea,liu09ea,yan09ea,qaz09ea} and quantum oscillation studies \cite{seb08ea, col08ea, ana09ea} in 1111 and 122 ferro pnictides of the form $Re$Fe$Pn$O and $Ae$Fe$_2$Pn$_2$ ($Re$ = rare earth, $Ae$ = alkaline earth, $Pn$ = As, P) consistently showed weak to moderate correlations with a sheet dependent mass enhancement $m^{*}/m_{\rm{band}}=1.3 - 2.1$. This is comparable to elemental Fe \cite{san09ea} and the phonon induced enhancement in MgB$_2$ \cite{yel02ea} but significantly lower that the renormalization seen in most transition metal oxides including the cuprate high temperature superconductors \cite{zho03ea,qaz09ea}. More recent systematic DMFT calculations reproduced this behavior and showed that earlier reports overestimated correlations mainly because of an incomplete account of the hybridization between Fe $d$ and pnictogen $p$ states \cite{aic09ea}.

In this letter we show that the above picture of weak correlations does not apply to the 11 family of iron based superconductors \festx. 
Our ARPES data from the ternary iron chalcogenide \fest{} show a simple non-reconstructed Fermi surface (FS) that is well reproduced by band structure calculations, which allows us to determine the mass enhancement for all Fermi surface sheets. In stark contrast to the moderately correlated (FePn)$^{-}$ families we find mass renormalization factors up to $m^{*}/m_{\rm{band}}\approx 20$ comparable to the most strongly enhanced transition metal oxides and exceeding the renormalization in cuprates by several times. Our findings are fully consistent with the high bulk specific heat coefficient of $\approx 39$~mJ/mol K$^{2}$ \cite{sal09ea} and demonstrate that the electronic properties of Fe(Se,Te) are remarkably different from the 122 and 1111 ferro-pnicitides. 

Iron--chalcogenides share the main structural motif of square planar sheets of tetrahedrally coordinated Fe with the 1111 and 122 (FePn)$^{-}$ families. Superconductivity with $T_c$ around 9 - 14~K and up to 37~K under moderate pressure is observed for a range of intermediate Se concentrations and for the end member Fe$_{1+\delta}$Se \cite{hsu08ea,mar09ea,gre09ea,sal09ea,che09ea,usu09}. DFT calculations indicate an electronic structure very similar to the ferro pnictides \cite{sub08,zha09} consistent with a recent ARPES study of non-superconducting Fe$_{1+\delta}$Te \cite{xia09ea}. Fe(Se,Te) does not order magnetically for intermediate Te concentrations but exhibits strong spin fluctuations with a propagation vector near $(\pi,\pi)$, where a static spin-density wave forms in iron-arsenides \cite{qiu09ea,lum09ea,moo09ea}. All these similarities point to the importance of the '11' compounds for understanding superconductivity in Fe based compounds.

The superconducting single crystals used in this work had T$_{c}=11.5(1)$~K and were grown in KCl/NaCl (1:1) flux at 800$^{\circ}$C from a powder precursor, which was prepared as described elsewhere \cite{gre09ea}. A composition Fe$_{1.0}$Se$_{0.42(2)}$Te$_{0.58(2)}$ was determined from single crystal x-ray diffraction and confirmed by energy dispersive x-ray analysis (EDX). ARPES experiments have been performed with a SPECS Phoibos 225 analyzer and a monochromatized He discharge lamp. Energy and angular resolutions were set to 12~meV / 0.3$^{\circ}$. All samples were cleaved at T $\approx12$~K along the $ab$-plane and measured at a pressure $<5\times10^{-11}$~mbar. Additional data at higher photon energies were taken at the SIS beamline of the Swiss Light Source.
Band structure calculation were performed for stoichiometric FeSe and FeTe using the augmented plane wave plus local orbital method implemented in the WIEN2K code \cite{bla2k} and the orthorhombic lattice parameters of Fe(Se,Te) from Gresty \textit{et al.} \cite{gre09ea} with relaxed chalcogen heights. The resulting electronic structure closely resembles the one reported in Ref. \cite{sub08} for a tetragonal unit cell.

\begin{figure}[tb]
\includegraphics[width=0.45\textwidth]{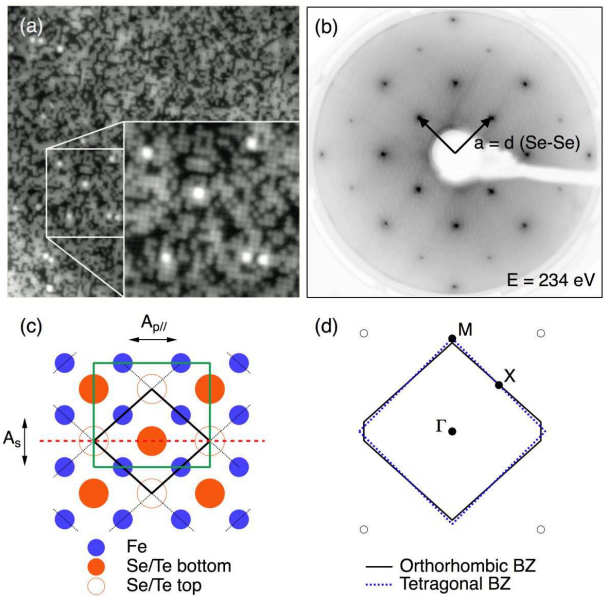}
\caption{\label{f1} (a) STM topograph of a cleaved surface taken at $T=6$~K in a 40~nm--square field of view with a 12~nm square inset showing the atomically resolved (Se,Te) surface layer ($V=146$~mV, $I=0.44$~nA). (b) LEED pattern taken with a primary beam energy of 234~eV. The reciprocal lattice vectors of the orthorhombic phase are indicated by black arrows. (c,d) Schematic of the crystal structure and Brillouin zone (not to scale). The conventional orthorhombic and a primitive unit cell of the low temperature orthorhombic phase are indicated in green and black in (c).}
\end{figure} 
%

% Fig 1: structure
In Fig.~1 we summarize the structural properties relevant for this paper. 
\fest{} undergoes a structural phase transition around 40~K from tetragonal to a metrically orthorhombic phase (space group $Cmma$) at low temperature \cite{gre09ea}. The associated orthorhombic strain $(a-b)/(a+b)=9\times10^{-4}$ is smaller than in FeSe but comparable to doped 122 and 1111 ferro-pnicitides \cite{nan09ea}. Importantly, neutron scattering does not show signs of magnetic ordering down to 4~K \cite{qiu09ea,lum09ea,moo09ea}. Hence, the structural transition does not increase the basis set and does not lead to a back-folding of the electronic band structure, thus removing much of the complexity found in the spin-density wave phase of pnictides. 
Moreover, the \festx{} system consists of neutral Fe(Se,Te) units as compared to the singly charged (FePn)$^{-}$ units in pnictides and lacks guest ions or interleaved slabs in the van der Waals gap.
This leads to structurally simpler surfaces of cleaved samples and renders the iron chalcogenides particularly amenable to surface sensitive probes such as ARPES \cite{mas09}.
The high surface quality achieved on \fest{} is evident from the topographic STM image in Fig.~1(a) showing the bulk truncated and atomically resolved (Te,Se) lattice over a large area. We identify the depressions with Se, the atoms appearing at intermediate height with Te and the occasional bright protrusions with excess iron atoms embedded in the chalcogen layer. From an analysis of $\approx 3600$ unit cells we derive a Se/Te content of 0.44(1)/0.56(1) and $\approx 0.1\%$ excess iron in excellent agreement with the EDX analysis.
The  absence of super--lattice spots and the relatively low background in low energy electron diffraction (LEED) patterns (Fig.~1(b)) confirms the above finding of a highly ordered surface with the translational symmetry of the bulk.

% Fig 2: establish the basic electronic structure
%
\begin{figure}[tb]
\includegraphics[width=0.48\textwidth]{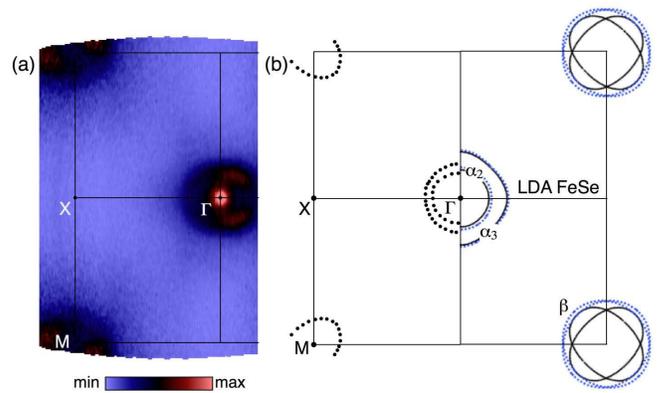}
\caption{\label{f2} (a) Fermi surface mapping of \fest{} measured with 21.2~eV excitation energy and $p$ polarization at T=12~K. The photoemission intensity has been integrated over 10 meV about the chemical potential. At this photon energy, $k_{z}$ is approximately 2.7 \AA, which is near the Z point in the third BZ. In (b) we compare the extracted contours with the DFT Fermi surface of FeSe. Dotted and solid lines correspond to the basal plane and mid-plane FS, respectively.}
\end{figure} 

The ARPES Fermi surface of \fest{} is shown in Fig.~2(a). The most intense feature at the $\Gamma$ point is due to a hole like band that barely touches the Fermi level as will be discussed later. Two weaker circular contours centered at the $\Gamma$ point define the $\alpha_{2,3}$ hole--like Fermi surface pockets with volumes of 2.5(5) and 4.1(8)\% of the Brillouin zone (BZ), respectively \cite{fn1}. This charge is compensated by slightly elongated electron pockets ($\beta$) at the M--points with an area of 3.2(7)\% BZ. 
Strong matrix element effects highlight the pocket with its long axis pointing radially away from the first $\Gamma$ point, while the second pocket is not discernible in our data \cite{fn2}.
A similar effect has been observed in the paramagnetic phase of Ba(Fe$_{1-x}$Co$_x$)As$_2$ \cite{liu09ea}.
Assuming two electron pockets of identical size and a two-dimensional Fermi surface we obtain a Luttinger volume of $-0.004\pm0.01$~electrons/unit cell, in excellent agreement with the expectation for a compensated metal.
The DFT calculation for FeSe shown in Fig.~2(b) qualitatively reproduces the experimental FS although it slightly overestimates the volumes. 
%As discussed in more detail in Fig.~4, the data agree poorly with calculations for FeTe, which predict three hole pockets at $\Gamma$.
Since the orthorhombic strain in \fest{} is over an order of magnitude smaller than shown in the sketch in Fig.~1(d) it has a minute effect on the shape of the Brillouin zone. For simplicity we thus plot all data in the square tetragonal zone.

% Fig 3: show the orbital dependent renormalization
%
\begin{figure}[tb]
\includegraphics[width=0.48\textwidth]{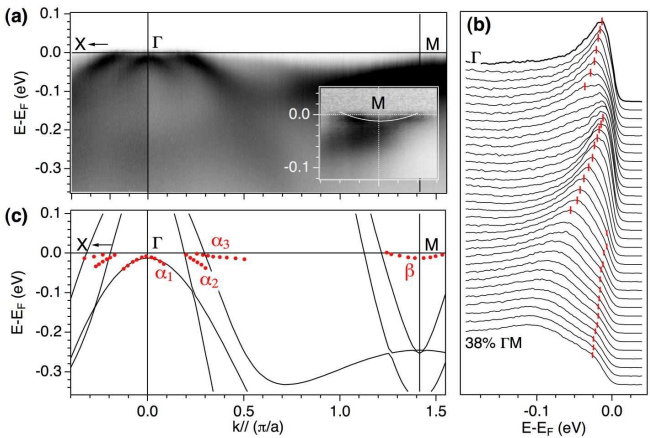}
\caption{\label{f3} (a,b) Band dispersion along X$\Gamma$M (21.2 eV, $s$--polarization, T=12~K). The inset shows data around the M--point, divided by a smooth function to enhance the contrast ($h\nu=36$~eV). The white line is a guide to the eye. (c) DFT band structure calculation for FeSe. The dispersion of the low-energy excitations extracted from the data in (a) is overlaid.}
\end{figure} 

The relatively simple Fermi surface of \fest{} has motivated us to investigate the interaction induced mass enhancement by comparing the measured dispersion with band structure calculations. Fig.~3(a) shows the ARPES intensity along X$\Gamma$M. 
At the $\Gamma$ point two intense hole-like bands are clearly resolved and a third weaker band is discernible. Although the $\alpha_1$ band contributes high intensity to the FS map (Fig.~2(a)) it does not contribute to the Fermi surface but reaches a top near -15~meV at the $\Gamma$ point. 
%In order to better visualize the M--point pocket seen in the FS map we normalized all MDCs around M to unit intensity and re-plotted the such processed data in the inset to panel (a). This procedure does not change the MDC derived dispersion but it causes small peak-shifts in EDCs, which we estimated numerically and corrected the data accordingly.
The shallow M-point electron pocket is more clearly visible in the inset showing data at $h\nu=36$~eV.

At very low energy we find a one-to-one correspondence between the quasiparticle excitations and the DFT band structure for FeSe shown in Fig.~3(b). However, the group velocities are strongly renormalized in the experiment. The effect is most marked for the $\alpha_3$ and $\beta$ sheets, which also have the lowest quasiparticle weights. 
We have extracted quasiparticle velocities from fits to energy distribution curves from multiple cleaves using empirical spectral functions and by analyzing the second derivatives.
For the $\alpha_3$ hole pocket, we obtain a Fermi velocity  $v_{F}\approx$~0.09 eV\AA{} corresponding to a renormalization of $m^{*}/m_{\rm{band}}=v_{\rm{band}}/v_{F}\approx17$. A comparable mass enhancement of $v_{\rm{band}}/v_{F}\approx20$ is found for the $\beta$ electron pocket, while the smaller $\alpha_2$ hole pocket with $v_{\rm{band}}/v_{F}\approx6$ is slightly less affected by interactions. Only the $\alpha_1$ band, which does not contribute to the Fermi surface retains a relatively high group velocity comparable to the calculation. 
%We note that all higher energy excitations seen in our data are strongly incoherent, which is consistent with the strong renormalization at the Fermi surface and in line with ARPES data from other correlated multi band systems showing similar mass enhancements \cite{tam08ea}.

% Fig 4: orbital character
%
\begin{figure}[tb]
\includegraphics[width=0.48\textwidth]{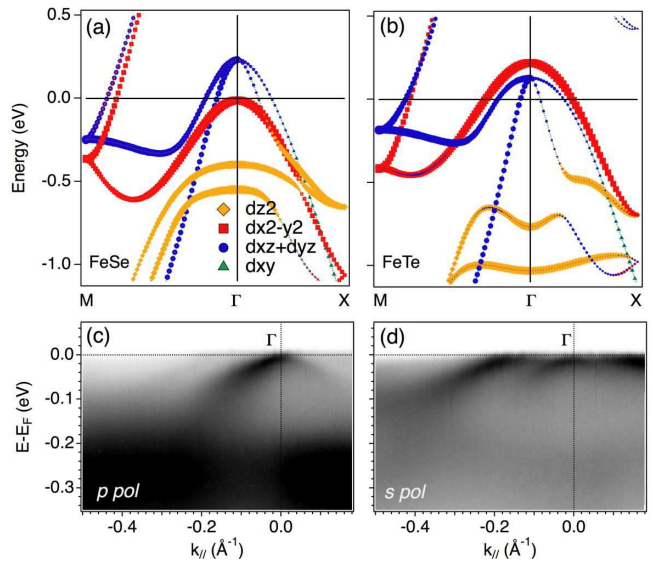}
\caption{\label{f4} (a,b) Orbital character of the DFT wave functions for FeSe and FeTe in a coordinate system defined by the Fe lattice. (c,d) Polarization dependence of the ARPES intensity along the $\Gamma$M mirror plane (see Fig.~1(c) for a definition of the scattering planes).  }
\end{figure} 
%
% s-pol is odd and suppresses even (x2-y2,z2,xz)
Use of the DFT band structure for FeSe to estimate the mass enhancement is supported by the polarization dependent ARPES measurements and calculations of the orbital character shown in Fig.~4. Strikingly, the calculations predict a different order of bands at the $\Gamma$ point for FeSe and FeTe, with the $d_{x^2-y^2}$ orbital below the $d_{xz,yz}$ bands in FeSe but above in FeTe. These orbitals can be distinguished in ARPES from the polarization dependence of their matrix elements.
Using dominantly odd polarization ($s$) with respect to the mirror plane indicated by the red dashed line in Fig.~1(c) suppresses orbitals of even symmetry ($d_{x^2-y^2}$, $d_{z^2}$, $d_{xz}$). Hence we can relate the bands with higher intensity in Fig.~4(c) ($p$--pol.) than (d) ($s$--pol.) to even orbitals. 
This strongly suggest that the $\alpha_1$ band is of $d_{x^2-y^2}$ character while the strong feature at -0.3~eV stems from $d_{z^2}$ bands, a behavior that is well reproduced by the calculation for FeSe but inconsistent with the order of bands predicted for FeTe.
%

% bulk sensitivity and composition in calculations
Before proceeding to a discussion of the above results we briefly consider possible shortcomings of our analysis. Given that DFT calculations indicate a polar surface for Fe(Te,Se) \cite{sub08} it is not \textit{a priori} clear that the near surface electronic structure measured by ARPES is bulk representative. For a metal, this can be tested stringently by comparing the electronic specific heat calculated from the sum of all low-energy excitations seen in ARPES with the direct measurement \cite{bau06bea}. To this end, we first calculate the quasiparticle masses $m^*=\hbar k_F/v_F$ using Fermi velocities and wave numbers averaged along $\Gamma$M and $\Gamma$X. This yields masses of 3.0(5), 16(5) and 11(4)~m$_e$ for $\alpha_2$, $\alpha_3$ and $\beta$, respectively, corresponding to a Sommerfeld coefficient of 29(6)~mJ/molK$^2$. The fair agreement with the direct measurement of 39 mJ/molK$^2$ \cite{sal09ea} rules out substantial errors in the experimental Fermi velocities.
We note that a previous ARPES study on non-superconducting Fe$_{1+\delta}$Te by Xia \textit{et al.} \cite{xia09ea} reported a much lower renormalization with $m^*/m_{\rm{band}}\approx 2$. However, this value corresponds to a Sommerfeld coefficient of $\approx 9$~mJ/molK$^2$, which is nearly a factor of 4 below the direct measurement \cite{che09ea} indicating that Ref.~\cite{xia09ea} underestimates the renormalization in the bulk.

%We next discuss a possible microscopic origin for the strong mass enhancement in \fest. 
Clearly, a sheet dependent mass enhancement between 6 - 20 is too high to be caused by electron--phonon interaction or the coupling to spin-fluctuations alone.  Instead our findings point to a dominant role of electronic correlations in the low-energy excitations of \fest. We note that DMFT calculations for ferro--pnictides applying correlations to the Fe $d$ shell only found much lower quasiparticle weights than observed experimentally \cite{hau08,aic09ea}. This suggests that subtle differences in the $p-d$ hybridization between the (FePn)$^{-}$ and Fe(Te,Se) families may account for the markedly higher renormalization in the chalcogenides  \cite{cra09, miy09}.
Interestingly, our DFT calculations show that the $\alpha_2$ sheet, which shows the lowest renormalization, has a higher chalcogen $p$ contribution to the wave function than $\alpha_{3}$.
Although correlations appear to dominate the mass enhancement in \fest, our data do not exclude a significant contribution from coupling to the strong near--$(\pi,\pi)$ spin fluctuations observed by neutron scattering \cite{qiu09ea,lum09ea,moo09ea}. In particular, coupling of the $\alpha_{2,3}$ sheets -- which are nested with the $\beta$ electron-pockets at the M--point -- might contribute to the higher renormalization of these sheets as compared to the $\alpha_1$ band. Though in the absence of a clear energy scale in the data such an interpretation remains speculative. We also point out that the strong correlations in Fe(Se,Te) lower the crossover energy between coherent quasiparticle states and strongly incoherent excitations with a more local character from several tenth of an eV in (FePn)$^{-}$ systems to 50 - 80~meV in Fe(Se,Te), comparable to other strongly correlated multi-band systems with a similar mass enhancement \cite{tam08ea}. This suggests that an itinerant interpretation of spin-fluctuations in iron chalcogenides is appropriate at low energy only.

% conclusions
In summary we have observed a strongly renormalized quasiparticle band structure in \fest{} with effective carrier masses up to 16~m$_{e}$. The mass enhancement is sheet dependent and roughly an order of magnitude larger than in the 1111 and 122 pnictides families. This demonstrates that the normal state of \fest{} is a strongly correlated metal and differs significantly from the ferro-pnictides. 
We hope that these findings motivate further theoretical work investigating the microscopic reasons for the strong mass enhancement and the remarkable insensitivity of superconductivity to the sizable change in the strength of correlations between \fest{} and the iron-arsenides.

% If you have acknowledgments, this puts in the proper section head.
\begin{acknowledgments}
We acknowledge discussions with C. Hicks and A.P. Mackenzie.
This work has been supported by the Scottish Funding Council, the European Research Council and the UK EPSRC (EP/C511794, EP/F006640). 
Part of this work was performed at the Swiss Light Source, Paul Scherrer Institut,  Switzerland.
%Crystal growth and characterization was supported by ...
\end{acknowledgments}

% Create the reference section using BibTeX:
%\bibliography{FBbib}

\begin{thebibliography}{36}
\bibitem{fn1}{Throughout this paper errors are estimated from the scatter of results obtained from different cleaves and by different fitting procedures.}
\bibitem{fn2}{Note that spin-orbit coupling lifts the degeneracy of the $\beta$-pockets along MX, resulting in two inequivalent pockets \cite{col08ea}. However, this effect is beyond the resolution of the present experiment.}
\end{thebibliography}

\end{document}